\documentclass[aps,prb]{revtex4b5}
\usepackage{epsfig}
\begin{document}
%
%\draft
%
\title{Slow fracton modes in antiferromagnetic correlations of high-T$_{c}$
superconductors}
\author{Mladen Prester}
 \email{prester@ifs.hr}
\affiliation{Institute of Physics, Zagreb, P.O.B. 304,
HR-10 000, Croatia}
\date{December 7, 1999}
\begin{abstract}
{In cuprate superconductors superconductivity develops as a unique crossover
between the two extremes characterizing these compounds, an antiferromagnetic
(AF) Mott insulator phase, on one side, and a non-superconducting itinerant metal
on the other.  The hole doping tunes a cuprate inside such a broad spectrum of
phases.  Numerous inelastic neutron scattering studies on underdoped samples
shows most directly that the static AF order of the parent phase does not
disappear by doping but transforms in such a way that at least fluctuating and
local AF order persists, coexisting with superconductivity.  Structurally, the
model of stripe-type correlations \cite{tra} accounts for the coexistence while
their genuine dynamics might hide \cite{kiv,cas} the most interesting physics of
cuprates, including the pairing mechanism.  Indeed, there is an impressive
accumulation of knowledge in favour of AF correlations underlying the pairing.
Some crucial properties of these correlations have recently been pointed out:
they should involve small velocity collective modes \cite{bal1} of the wave
vector identical to that of the static AF order \cite{cam} while their energy
scale \cite{bal1,car} ($<50 meV$) should be rather low compared with any other
involved \cite{bal2}.  Here we show on phenomenological ground that the latter
dynamics originates, irrespective of structural details of stripe order, from
general geometrical constraints known to characterize the problem of disordered
interpenetrated phases.  In particular, we show that dynamics of the
Euclidean-to-fractal crossover introduces a new (10-50 meV) energy scale,
associated with slow (or almost localized) and well-defined fracton modes, and
provides a consistent and natural interpretation for redistribution of spectral
weight and other inelastic neutron scattering observations.  We conclude
therefore that the fracton excitations substantially contribute to pairing
interactions in high-T$_{c}$ cuprates.}
\end{abstract} 
\maketitle
\section*{}
High-resolution studies of dissipation in weak link network of polycrystalline
cuprates have recently identified a fractal dissipative regime inside which a
self-organized current-carrying medium is an object of fractal geometry
\cite{pre1}.  Taking into account that the Josephson-coupled disordered medium
represents a good description of the problem of phase coherence not only {\em
between} the CuO$_{2}$ planes but also {\em inside} them (due to stripes
and/or other forms of charge separation and clustering, filamentary
fragmentation, etc.)  the presence of fractal geometry in self-organized
structures of CuO$_{2}$ planes seems rather plausible as well.  In this work
we investigate elastical modes in such a hypothetical (but still physically
justified) two dimensional (2D) fractal lattice.  In particular, we study AF
spin fluctuations inside a model which allows a fractal geometry of the
relevant AF cluster; the latter, being sample-sized and homogeneous in the
parent compound, gets progressively more ramified by hole doping.  (The
liquid-crystal-like stripe phases \cite{kiv} represent of course a closely
related visualization).

Generally, the dynamical features of such a problem are well-known:  instead
of extended modes (phonons or magnons) of elastic quasi-continuum the
vibrational modes in fractal lattices are localized excitations, known as
fractons \cite{orb,nak1}.  The fractality of real lattices depends on wave
vector of the considered excitation so the fracton modes start to dominate
only above some critical wave vector, i.e., above the frequency of
phonon-to-fracton crossover \cite{nak1}.  Also, there is a considerable
accumulated knowledge about fractons specifically in classical diluted
antiferromagnets both in their theoretical \cite{yak} and experimental
\cite{uem,ike} aspects.  In these classical systems the magnetic component
responsible for static AF order (e.g., Mn$^{2+}$ ion in MnF$_{2}$ or in
RbMnF$_{3}$) is substituted at random by non-magnetic substituent (e.g.,
Zn$^{2+}$ or Mg$^{2+}$, respectively).  Inelastic neutron scattering studies
on samples comprising substituent in concentrations approaching threshold
concentration (above which there would be no magnetic order) revealed
dynamical features consistent with predictions for AF fractons \cite{uem,ike}.
However, besides some doubts regarding proper assignment of specific
experimental features \cite{tak} there are still some uncertainties about the
proper analytical form of the dynamical structure factor to be applied in
modeling fractal antiferromagnets.

Applying this knowledge to the problem of underdoped cuprates we first
simplify our elaboration by limiting ourselves to the case of localized
magnetic moments (spins associated to Cu$^{2+}$ ions) only.  There is a
consensus now that this is entirely appropriate for the spin wave phase of
parent AF compounds but also equally inappropriate for the overdoped
non-superconducting systems being compatible with itinerant electron
description.  The present work explores primarily the underdoped range.
Localized description certainly makes sense for low hole concentration but
becomes questionable close to optimal doping.  There is however an increasing
number of arguments \cite{lee} pointing out the localized nature of
incommensurate magnetism even in optimally superconducting samples (oxygen
doped La$_{2}$CuO$_{4+y}$) so one could expect that the model we elaborate
here may be applied at least to a reasonably wide range of underdoped
concentrations.  Now we define our `building block' which, playing the role of
a non-magnetic ion in diluted antiferromagnets, are to be placed at random in
the initially homogeneous and antiferromagnetic CuO$_{2}$ plane.  The
fluctuating stripe correlations are certainly the most convincing
interpretation for the incommensurate magnetism in superconducting cuprates
and the local stripe order seems to be an inevitable element of cuprate
physics.  Our building blocks are therefore the regions with stripe order,
disordered in sizes and, possibly, in other features.  By `stripe order' we
mean, in accordance with the localized magnetism of the present model, the
original \cite{tra} pattern of `the three spins wide AF stripes separated by
non-magnetic antiphase domain boundary' \cite{tra}.  The internal structure of
the building block complicates somewhat the geometrical percolation scenario
of the diluted antiferromagnets; thus, we do not attempt to define the
threshold hole concentration here.  However, one expects that the main
qualitative elements of an crossover to fracton dynamics, in increasing hole
concentration, remain valid.

Qualitatively, the scenario for underdoped cuprates would be the following:  With no holes added
the magnetic collective modes are high-velocity ($\hbar v=650meV \AA$, typically) regularly damped
spin waves (Fig.1a) characterized by dominantly 2D dynamics.  The latter follows from the very high
inplane superexchange (100meV, typically) compared with the interplane one.  Added holes introduce
random obstacles for spin waves localizing them into fracton modes at short wave lengths
($q>q_{c}$), i.e., above the hole-concentration-dependent cross over energy $\hbar \omega_{c}=\hbar
vq_{c}$.  This takes place when ramified AF cluster becomes self-similar (fractal) on all spatial
scales smaller than $1/q_{c}$; subsequently, the crossover frequency progressively decreases as
hole concentration (thus cluster fractality) increases \cite{nak1}.

Quantitatively, all of these dependencies must obey specific power-laws
involving several statical and dynamical exponents \cite{nak1,orb} (e.g.,
correlation length and dynamical one, fractal and AF fracton dimensions) which
are, at least approximately, known \cite{nak1}.  The dynamical structure
factor $S(q,\omega)$ has to include these power-laws and to assure known
asymptotic behaviour characterizing the homogeneous ($\omega \ll \omega_{c}$)
and the fractal ($\omega \gg \omega_{c}$) regime.  The problem of dynamical
structure factor for AF fractons has been treated by many approaches
\cite{nak1} most of which are founded on the effective-medium considerations
\cite{yu,chr}.  Out of several probably equivalent results for $S(q,\omega)$
in this work we use the form of Polatsek and Entin-Wohlman \cite{pol} as it
explicitly demonstrates the localization features of fracton modes (Fig.1b).
Although the latter result suffers from general shortcomings of all effective
medium approaches, why it cannot quantitatively be correct in all details
\cite{nak1}, its general Green's function background assures its basic
usefulness.  This form has been successfully used in studies of fractons in
silica aerogels \cite{cou}, and, more recently, in fused silica \cite{for}.
Therefore, the form we propose to determine the scattering cross section of AF
spin waves in cuprate CuO$_{2}$ planes of increasing fractality reads as

\begin{equation}
S(q,\omega)= \Bigl[\frac{v(\omega)q}{\omega^2}\Bigr]\Bigl[
\frac{\Gamma(\omega)}{(\omega-v(\omega)q)^2+\Gamma^2(\omega)} 
-\frac{\Gamma(\omega)}{(\omega+v(\omega)q)^2+\Gamma^2(\omega)}\Bigr],
\end{equation}
where $v(\omega)$ and $\Gamma(\omega)$ are velocity of spin waves and damping
(lifetime) function, respectively.  The frequency dependence of these two
quantities makes the present $S(q,\omega)$ different, together with its
prefactor, from standard Lorentzian `shape functions' generally used in
scattering problems and, particularly, in the inelastic neutron studies on
diluted antiferromagnets \cite{coo}.  The asymptotic forms of $v(\omega)$ and
$\Gamma(\omega)$ are precisely known \cite{nak1}:  At low $\omega$
$(\omega<\omega_{c})$, $v=v_{0}$, the magnon velocity, while at high $\omega$
$(\omega>\omega_{c})$, in fractal regime, it must obey $v(\omega) \propto
\omega ^{z}$ (z is dynamical exponent).  In the same limits the damping
$\Gamma(\omega)$ crosses over from the phonon (magnon)-like Rayleigh law (
$\Gamma(\omega) \propto \omega ^{d+1}/ \omega_{c} ^{d}$, d is Euclidean
dimension) to the localized Ioffe-Regel ($\Gamma(\omega) \propto \omega$)
regime \cite{aha} where both the dispersion and the wave vector description
becomes ill-defined.  For the specific frequency dependence of $v$ and $
\Gamma $ we use the heuristic expressions of Courtens et.al \cite{cou} shown
to work well in different fractal systems \cite{for}.  Taking into account a
predominantly 2D nature of magnetism in CuO$_{2}$ planes the expressions are:

\begin{equation} 
%\begin{array}{l}
v(\omega ) = v_0 \left[ {1 + (\omega /\omega _c )^m } \right]^{z/m} 
\hspace {.5cm}, \hspace {.5cm}
\Gamma(\omega )= \Gamma _0 \omega _c (\omega /\omega _c )^3 \left[ {1 + (\omega /\omega _c
)^m } \right]^{ - 2/m},
\end{equation}
where $m$ characterizes the sharpness of the crossover.  Most of the features
of the present model have a weak dependence on the choice of $m$ and the
`traditional' \cite{cou,for} value $m=4$ has been used.  Also, in calculation
of $z=d_{f}/ \tilde{d}_{af}$, where $d_{f}$ is the fractal and
$\tilde{d}_{af}$ the AF fracton dimension, a well-accepted\cite{nak1} values
$d_{f} \approx 1.9$ and $\tilde{d}_{af} \approx 0.9$ have been used.  We note
however that our results depend very little, at least qualitatively, on the
fine tuning of exponents.  Fig.1 visualizes $S(q,\omega)$ of Eq.1 taking into
account real physical parameters of YBCO superconductor.

Now we compare $S(q,\omega)$ with the experimental results for imaginary part
of generalized susceptibility $\chi^{"}({\bf q},\omega)$ on cuprates which
determines the scattering cross section.  These results are summarized in,
e.g., Refs.  \onlinecite{hay,fon}.
%figure1
\begin{figure}[ht]
\epsfig{file=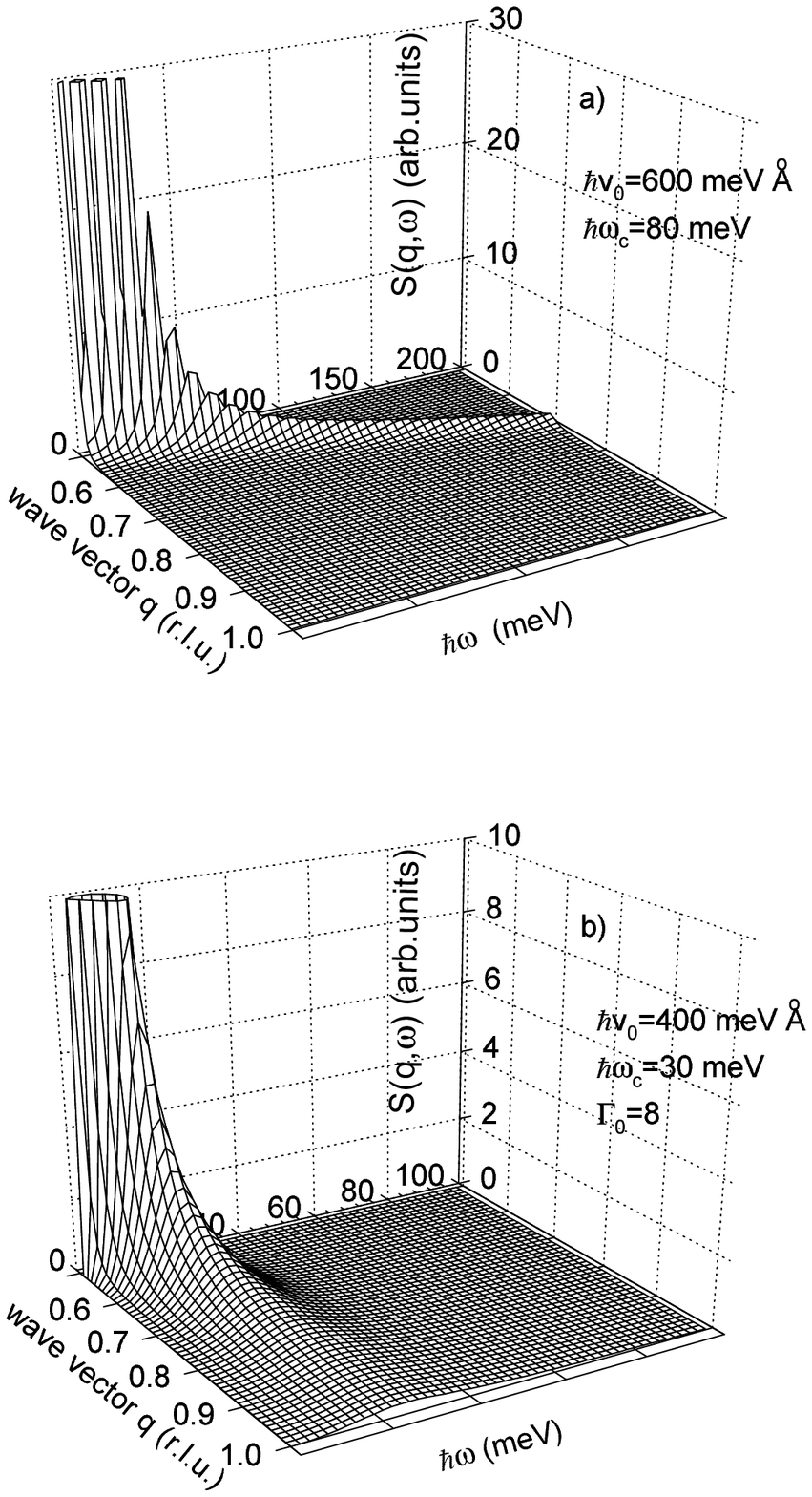,height=19cm,clip=}
\caption{Dynamical structure factor $S(q,\omega)$ for inelastic neutron
scattering by cuprates' AF correlations of this model, Eqs.1,2.  The two
asymptotic dynamics is shown:  dispersive spin waves (shown in a)) and
localized fractons (shown in b)).  Lattice parameters of YBCO
superconductor have been assumed.  Known velocity of spin waves $v_{0}$
has been used as a parameter.  Some reduction of $v_{0}$ by hole doping
has been both observed and expected.  The form of Eq.2 has been used for
damping; however, in a) the damping was somewhat additionally suppressed
in order to visualize the spin waves.  The main physical parameter is
the crossover frequency $\omega_{c}$.  It decreases by hole doping and
defines the energy scale of localizing fracton modes.  The feature of
localization is reflected by vanishing wave vector dependence for
$\omega > \omega_{c}$, where the mode extension becomes parallel to
q-axis.}
\label{1}
\end{figure}
Results at fixed q ($\omega$-scan) or at
fixed $\omega$ (q-scan) depend very much on convolution with instrumental
resolution \cite{hay1} involving the integration of intrinsic $S(q,\omega)$
inside usually a sizable resolution ellipse.  Hence, a better defined
experimental quantities are momentum- or energy-integrated $\chi^{"}({\bf
q},\omega)$.  The local (Brillouin zone averaged) susceptibility
$\chi_{2D}^{"}(\omega) \propto \int \chi^{"}({\bf q},\omega)d^{2}q$ is
particularly well investigated \cite{hay,fon}.  In hole doped samples
$\chi_{2D}^{"}(\omega)$ shows a characteristic maximum (Fig.2b,c) in the range
of 20-50 meV while the parent compounds reveal approximately energy
independent $\chi_{2D}^{"}$ \cite{hay,fon}, in agreement with
$S^{SW}(q,\omega)$ for spin waves.  It is important to note that a sizable
area of $\chi_{2D}^{"}(\omega)$ peaks is positioned, in graphs involving
absolute calibration, {\em above} the spine wave level.  The latter reflects a
substantial redistribution of spectral weight in frequency domain as the holes
are added to the planes.  The present model for $S(q,\omega)$ readily accounts
for the redistribution.  The 2D integration of our model's $S(q,\omega)$,
Eq.1, gives:

\begin{equation} 
S_{2D}(\omega) \equiv \int\limits_{B.z.}  {S(q,\omega )d^2 q} = \frac{{2\pi \Gamma
}}{{\omega^2 v^2}}\biggl[ {\Gamma A(\omega) \Bigl( {\frac{{\omega ^2 }}{{\Gamma ^2 }} - 1}
\Bigr) 
+ \omega \ln \Bigl[ {\frac{{(v Q - \omega)^2 + \Gamma
^2 }}{{\omega ^2 + \Gamma ^2 }} \cdot \frac{{(v Q + \omega )^2 + \Gamma ^2
}}{{\omega ^2 + \Gamma ^2 }}} \Bigr]} \biggr]
\end{equation}
where Q is Brillouin zone boundary and 
\begin{equation} 
A(\omega)=arctg\frac{{v Q - \omega }}{\Gamma } - arctg\frac{{v Q
+ \omega }}{\Gamma } + 2arctg\frac{\omega }{\Gamma }.
\end{equation}

We note again that $v$ and $\Gamma$ are $\omega$-dependent (Eq.2).  The result
for $S_{2D}(\omega)$, which makes a central result of this paper, is shown in
Fig.2a for one particular choice of parameters together with the level which
would characterize magnons propagating with velocity $v_{0}$.  Except
behaviour just above the origin there is a full reproduction of experimental
curves.  The physical reason for the redistribution in our model is clear
immediately from Fig.1:  For very small energies the magnetic modes start to
propagate as magnons but soon, by approaching $\omega_{c}$, slow down and
accumulate in the region around $\omega_{c}$ taking away the spectral weight
that magnons would have in the hole-free lattice up to high energies.  We
discover therefore that $\hbar \omega_{c}$ represents a natural new scale of
energy characterizing the AF correlations in cuprates.  $S_{2D}(\omega)$ of
Eq.3 perfectly models most of the experimental results by choosing $\hbar
\omega_{c}$ from the interval $10-50 meV$.  The present model attributes the
experimental peaks in $\chi_{2D}^{"}$ (and also in $\chi^{"}(q,\omega)$, see
below) to dynamics of crossover quantified, in this work, by Eqs.2,3.  One
should note, however, that the main features are not determined by some
special choice of spectral function $S(q,\omega)$, such as the one of Eq.1;
instead, the features stem from universal principles of matching the two
different regimes characterized by different geometrical constraints.  In our
model there is no principal difference between the narrow `resonant' peak
\cite{bou}, setting-in below T$_{c}$, and the broad off-resonant
\cite{bou,bal1} peaks as they both reflect the dynamics of crossover.  The
experimental differences may come primarily from the involved parameters, such
as the damping scale $\Gamma_{0}$.  Indeed, it seems very plausible that the
macroscopic superconductivity of the optimum-doped samples allows higher level
of fractal self-organization accompanied with the increased value of
$\Gamma_{0}$.  One easily verifies that the increased $\Gamma_{0}$ induces, in
turn, a more pronounced peak in $S_{2D}(\omega)$.
%Figure 2
\begin{figure}
\epsfig{file=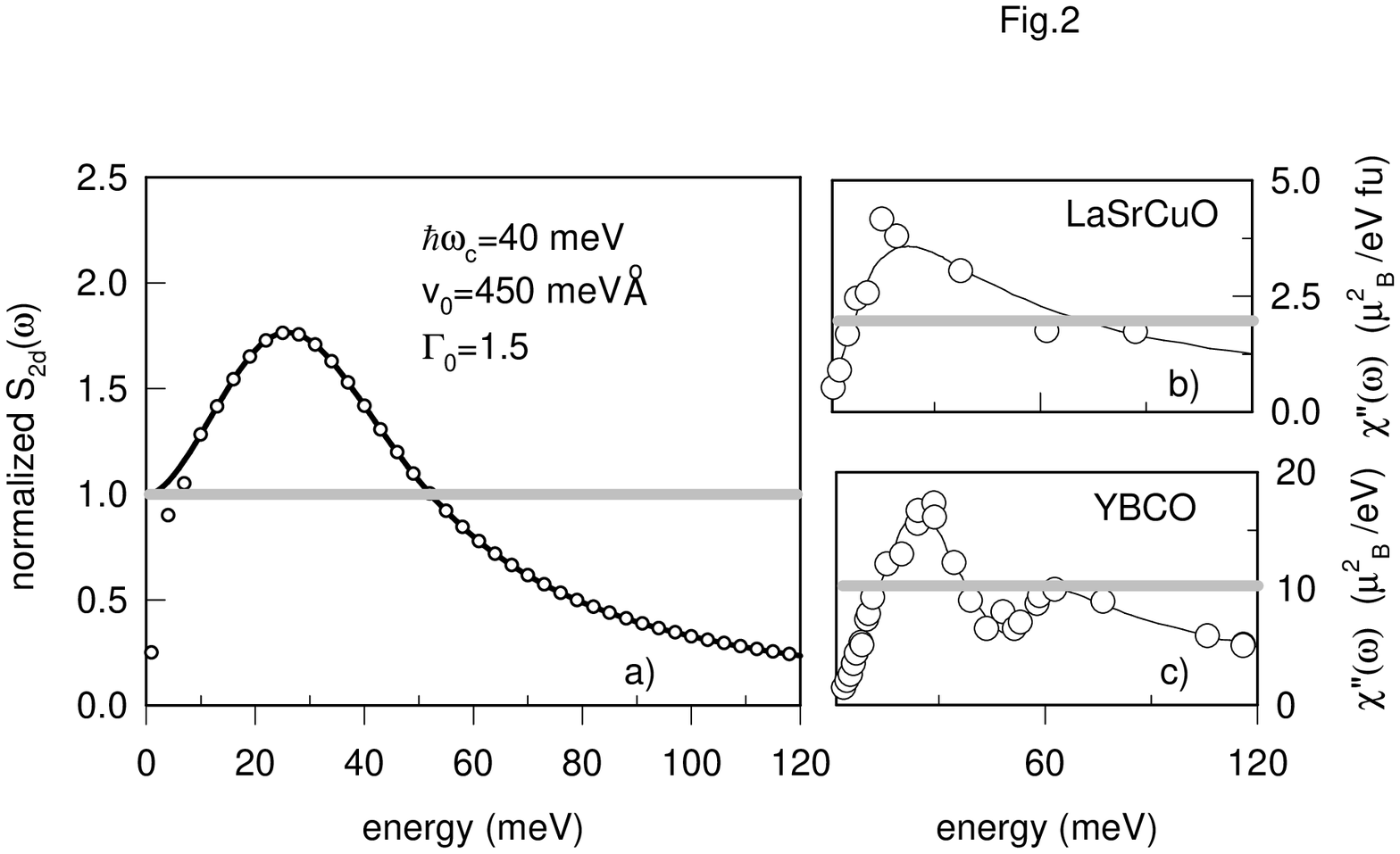,width=16cm,clip=}
\caption{Integrated (local) dynamical structure factor $S_{2D}(\omega)$
of Eqs.3,4, normalized to $S_{2D}$ of AF magnons characterizing the
Euclidean parent phase, is shown in a) (full black line).  A peak
develops as a result of accumulating modes with frequencies close to the
crossover ($\omega_{c}$) one.  Numerical integration of Eq.1 has also
been performed (open circles).  Disagreement for small energies
represent an artifact of numerical integration of very sharp functions.
Experimental results for local susceptibilities of doped superconducting
cuprates are shown in b), Ref.23, and c), Ref.24,26 on the same energy
scale.  Thick gray lines in all panels represent the level associated to
propagating spin waves.}
\label{2}
\end{figure} %

Now we analyze results for fixed fixed q ($\omega$-scans) or at fixed $\omega$ (q-scans) in the
light of our model, Fig.3.  In literature, there are many data \cite{bou} for $\chi^{"}$ with q
fixed at AF wave vector, ${\bf q}=(0.5,0.5)$ in reciprocal lattice units ($rlu$), measured on a
variety of underdoped and optimally doped samples (no significant magnetic correlations

%Figure 3
\begin{figure}
\epsfig{file=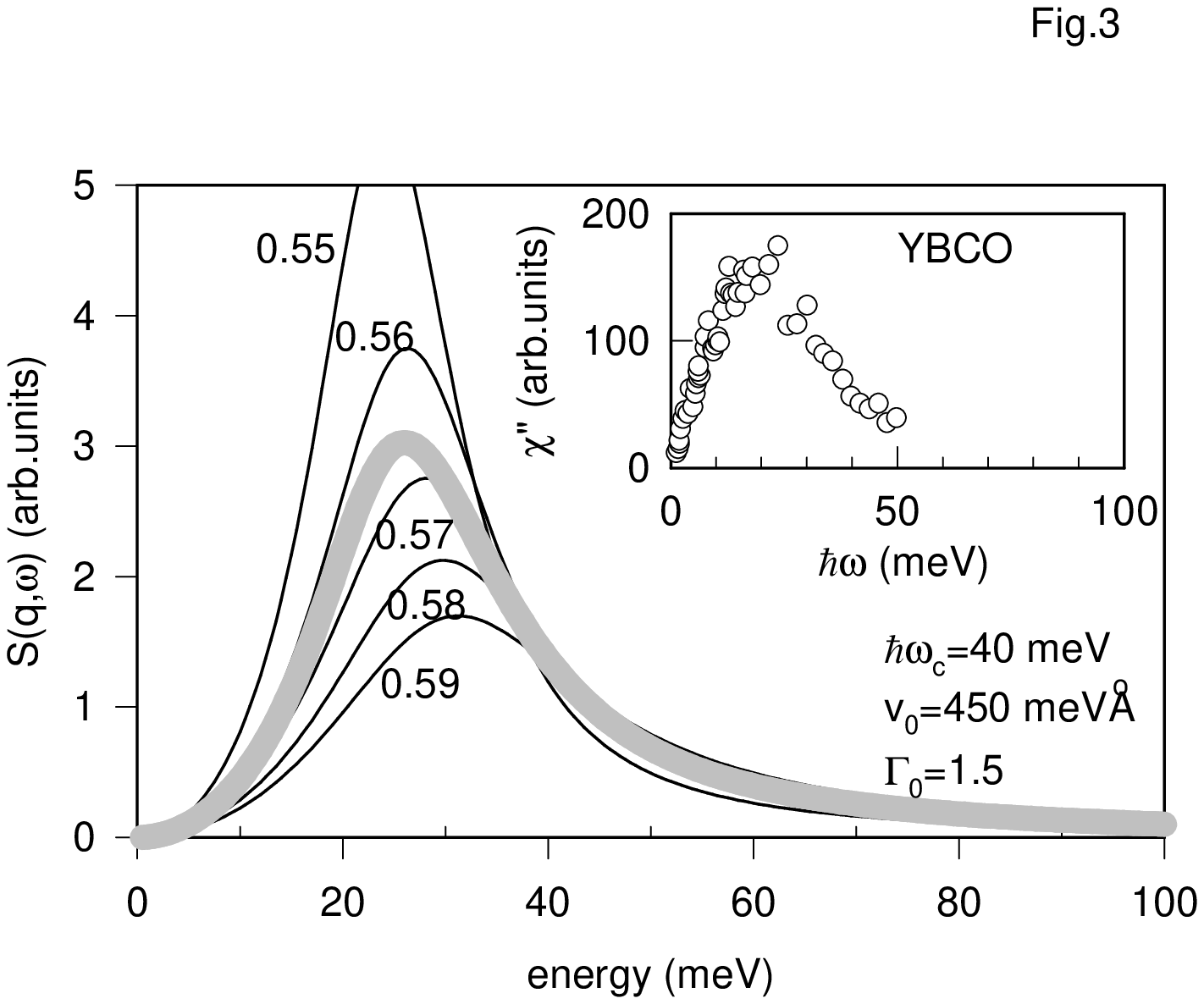,width=14cm,clip=}
\caption{$\omega$-dependence of $S(q,\omega)$, Eqs.1,2, for five fixed
wave vectors close to the AF zone center at $q_{AF}=0.5$.  The numbers
near the curves designate the parametric values of $q$ in reciprocal
lattice units $2\pi /a \propto 1.63 \AA ^{-1}$.  A crossover to a weak
$q$-dependence (i.e., localization) obviously takes place in this
interval of $q$.  Thick gray line is the average of curves.  Depending on
parameters, there are sharp propagating magnon peaks (not shown)
immediately above $q_{AF}$ ($0.5<q<0.55$, typically).  Inset:
experimental result on underdoped superconductor, Ref.26}
\end{figure}

has been observed in overdoped range).  These results reveal a pronounced peak
with the maximum positioned usually inside the interval 20-50 meV but with
other details (e.g., temperature dependence, energy width) being dependent on
the sample family and the extent of doping.  In particular, the resonant peak
remains as the only magnetic contribution to $\chi^{"}$ in optimally doped
samples (not yet observed in La$_{2-x}$Sr$_{x}$CuO$_{4}$ family).  In our
model the region of Brillouin zone which belongs to the close vicinity of
${\bf q}=(0.5,0.5)$ (say, for $0.5rlu<q<0.6rlu$, Fig.3) is subject to an
abrupt transformation of modes from a purely dispersive (spine wave) to almost
localized (fracton) ones.  A $\omega$-scan taken nominally at ${\bf
q}=(0.5,0.5)$ actually integrates the neighbouring region in which this
transformation takes place.  In the integration, the dispersive mode gives an
$\omega$-independent background contribution while the modes accumulating
close to $\omega_{c}$ contribute to the net $\omega$-dependence of
$\chi^{"}((0.5,0.5),\omega)$.  This is illustrated and further commented in
Fig.3.

Finally, we discuss the q-scans.  The neutron data have shown that there is a
characteristic linear relationship \cite{yam,bal1} between the q-space width
of $\chi^{"}(q,\omega)$ and T$_{c}$ at constant $\omega$ .  The slope of this
dependence, having the dimension of velocity \cite{bal1}, has been used
\cite{bal1,bal2} as an evidence for the slow excitation, perhaps with a
crucial contribution to formation of the superconducting state.  In our model,
a pronounced q-width of $\chi^{"}({\bf q},\omega)$ is a natural consequence of
progressive slowing down and localization of magnetic modes close to
$\omega_{c}$.  The peak q-width depends on the value of $\omega_{c}$
determined, in turn, by the hole concentration.  Fig.4 illustrates the
increase of the q-width by decreasing $\omega_{c}$ demonstrating that an
approximately linear relationship is well-obeyed in a broad range of crossover
frequencies.
%Figure 4
\begin{figure}
\epsfig{file=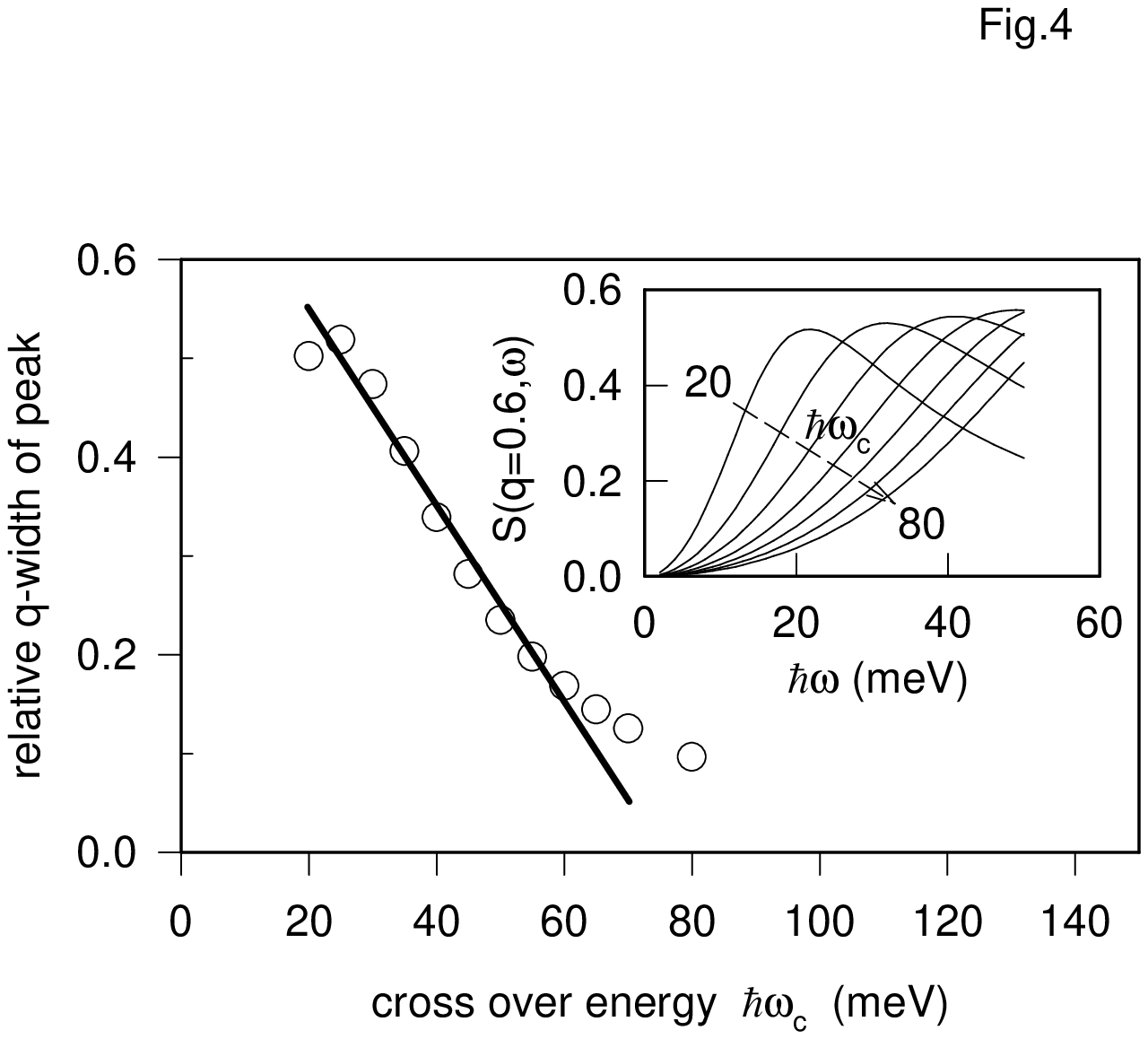,width=14cm,clip=}
\caption{Dependence of the relative $q$-width of $S(q,\omega)$ on
doping-dependent parameter $\omega_{c}$.  Inset shows a set of
$\omega$-scans of $S(q,\omega)$, Eqs.1,2, for $q$ fixed at $q=0.6$
differing only in the designated choice for $\omega_{c}$.  The amplitude
of $S(0.6,\omega)$, taken at $\hbar \omega =25meV$, has been chosen as a
relative measure for the $q$-width.  One notes a broad range of
$\omega_{c}$ characterized by approximately linear behaviour of
$q$-width.  Full line is a guide for the eye.}
\label{4}
\end{figure}

We conclude therefore that there is a high level of agreement between the
present model and the experimental results for imaginary susceptibility and
that the slow fracton modes could provide a basis for interpretation of the
neutron studies.  We also claim that these modes could provide a basis for
understanding pairing in high-T$_{c}$ superconductors.  Generally, the
fracton-mediated superconductivity has been treated theoretically in framework
of quasi-crystals \cite{mou} by solving an appropriate Eliashberg equation.
On other hand, the coupling of carriers to 40 meV `resonant spin excitations'
has recently been recognized \cite{car} as strong enough to account for high
values of T$_{c}$ in cuprates.  The present work demonstrates that the fracton
modes possibly underly these excitations and we believe that they could be a
clue for `uncovering the nature of slow moving charge objects'\cite{bal2},
thus providing `a crucial step for our understanding of high-T$_{c}$
superconductivity'\cite{bal2}.

%

%GdBa$_{2}$Cu$_{3}$O$_{7-x}$
%Fig.\ \ref{6}
%Eq.\ (\ref{e3})
%CC=\cite
%SQO=$S(q,\omega)$

%vv=$v(\omega)$
%gg=$\Gamma(\omega)$
%KK=$\chi^{"}({\bf q},\omega)$
%CC=\cite
%f_{B}
%K2=$\chi_{2D}^{"}(\omega)$
%S2D=$S_{2D}(\omega)$
%OC=$\omega_{c}$
%*=$\omega$
%

%

%
{\bf Acknowledgements.}Discussions with D.Pavuna, S.Tomi\'c, D.Djurek and A.\v Siber are
gratefully aknowledged.
\end{document}